\documentclass[12pt,preprint]{aastex}

\def\lap{\lower.5ex\hbox{$\; \buildrel < \over \sim \;$}}
\def\gap{\lower.5ex\hbox{$\; \buildrel > \over \sim \;$}}

\usepackage{natbib}
\begin{document}
\title{Doppler Shifts and Broadening and the
Structure of the X-ray Emission from Algol}

\author{Sun Mi Chung\altaffilmark{1}, Jeremy J. Drake\altaffilmark{1},
Vinay L. Kashyap\altaffilmark{1}, 
Li Wei Lin\altaffilmark{1}, Peter W. Ratzlaff\altaffilmark{1}}
\affil{Harvard-Smithsonian Center for Astrophysics, \\ 60 Garden Street, \\
  Cambridge, MA 02138}

\begin{abstract}
In a study of {\it
Chandra} High Energy Transmission Grating spectra of Algol, we clearly
detect Doppler shifts caused by the orbital motion of Algol~B.  These
data provide the first definitive proof that the X-ray emission of
Algol is dominated by the secondary, in concordance with expectations
that the primary B8 component should be X-ray dark.  However, the
measured Doppler shifts are slightly smaller than might be expected,
implying an effective orbital semi-major axis of about $10R_\odot$
instead of $11.5R_\odot$ for the Algol~B center of mass.  This could
be caused by a small contribution of Algol~A, possibly through
accretion, to the observed X-ray flux, in which case such a
contribution does not exceed 10-15\%.  We suggest the more likely
explanation is an asymmetric corona biased toward the system center of
mass by the tidal distortion of the surface of Algol~B.  A detailed
analysis of the profiles of the strongest lines indicates the presence
of excess line broadening amounting to approximately 150~km~s$^{-1}$
above that expected from thermal motion and surface rotation.
Possible explanations for this additional broadening include
turbulence, flows or explosive events, or rotational broadening from a
radially extended corona.  We favor the latter scenario and infer
that a significant component of the 
corona at temperatures $<10^7$~K has a scale height
of order the stellar radius.  This interpretation is supported by the
shape of the X-ray lightcurve and tentative detection of a shallow dip
at secondary eclipse.  We also examine the O~VII intercombination and
forbidden lines in a Low Energy Transmission Grating Spectrograph
observation and find no change in their relative line fluxes as the
system goes from quadrature to primary eclipse.  Since these lines
appear to be strongly affected by UV irradiation from Algol~A through
radiative excitation of the $2\,^3S\rightarrow 2\,^3P$ transition,
this supports the conjecture that the corona of Algol~B at
temperatures of several million K must be significantly extended
and/or located toward the poles to
avoid being shadowed from Algol~A during primary eclipse.

\end{abstract}

\keywords{stars: X-rays --- stars: binary --- stars: activity --- stars: linewidths  }

\section{Introduction}
\label{s:intro}

Algol ($\beta$ Perseus) is the eponymous eclipsing binary system which
consists of a primary early-type
main sequence star and a Roche lobe-filling secondary late-type giant
or subgiant star.  These systems have undergone a period of mass
transfer during which material was transferred from the initially more
massive present day late-type star to its initially less massive
early-type companion.  Algol is the brightest and nearest example of
this type.  
The primary star (Algol A) is a B8 V
main sequence star, while the secondary (Algol B) is a K2 IV
subgiant that has lost about half of its original mass to the present
day primary \citep[e.g.,][]{Drake2003}.
Since Algol has a short period orbit of
2.87 days \citep[e.g.,][]{Hill1971} the two stars are tidally locked, and
their orbital and rotational periods are synchronized.

Algol was first confirmed as an X-ray source by \citet{Harnden1977},
who suggested that the X-ray emission might be explained by a
mass-transfer model, where the more massive B star accretes material
from the less massive K star, either through Roche Lobe overflow or
a stellar wind.  In both cases, the infalling
material is shock-heated to X-ray temperatures when it hits the B star.   
According to this mass-transfer model, we would expect to observe an
X-ray eclipse during the optical primary eclipse.  However, 
observations made with the Solid State Spectrometer
(SSS) of the \emph{Einstein} Observatory revealed no sign of an X-ray
eclipse at the time of optical primary eclipse \citep{White1980}.

At present, it is widely believed that the X-ray emission from Algol
arises mostly, if not completely, from the corona of Algol B
\citep{White1980}.  The two stars are tidally locked and rapidly
rotating, and within the rotationally-excited dynamo paradigm, it is
expected that the convection zone of the K star would experience
increased dynamo activity whose resulting magnetic energy is
subsequently dissipated at the stellar surface in the form of a hot,
X-ray emitting corona \citep{Pallavicini1981,Simon1989}.  However, the
nature and structure of this corona remains a topic of debate.  More
than four orders of magnitude in X-ray luminosity and an order of
magnitude in plasma temperature separate the solar corona from coronae
of most active stars such as the Algol-type secondaries 
\citep[e.g.][]{Schmitt1997}.  
We currently have very little idea of the physical and spatial
characteristics and appearance of these much more active coronae, and
whether or not magnetic structures are restricted to the stellar
surface or whether significant X-ray emitting plasma resides in larger
structures, including inter-binary loops, with magnetic structures
linking the two stars \citep{Uchida1983}.

\citet{White1986} analyzed EXOSAT observations of Algol and found no
clear X-ray eclipse during the optical secondary minimum.  If this
X-ray emission was indeed from the secondary star, his study suggests
that the coronal extent of Algol B must be comparable to or greater
than the size of the star itself.  However, \citet{Ottmann1994} did
detect a secondary X-ray eclipse in ROSAT observations, and concluded
that the scale height of the Algol B corona is $0.8 R_B$.
Similarly, a shallow eclipse at optical secondary minimum was seen by
ASCA \citep{Antunes1994}.  A dramatic eclipse of a large flare
observed by BeppoSAX enabled the size of the flaring structure to be
estimated as being less than 0.6 stellar radii and pinpointed the
location of the structure to polar regions \citep{Schmitt1999}.
Based on these deductions and on observations of flares on Algol seen
by GINGA, EXOSAT, ROSAT, and XMM-Newton, \citet{Favata2000} suggest that the
corona is essentially concentrated onto the polar regions of the K
star, with a more compact (smaller than the star) flaring component
and a perhaps somewhat more extended (comparable {in size} to the
star) quiescent corona.  Such a picture of polar-dominated activity
has long been suspected based on optical Doppler imaging techniques,
which consistently see dark polar spots on active stars
\citep[e.g.][]{Strassmeier2000}.  \citet{Ness2002}
also suggested
polar emission was responsible for the lack of rotational modulation
in the observed flux in a 
{\it Chandra} Low Energy Transmission Grating Spectrograph (LETGS)
observation of Algol.  However, an XMM-Newton observation of an Algol flare
studied by \citet{Schmitt2003_XMM} was interpreted as lying at lower
latitudes.  While this interpretation is somewhat subjective, it does
suggest that even if polar emission dominates, significant activity at
all latitudes is likely present. 
 
Similar conclusions regarding the corona of the K star might be drawn
based on radio detections of Algol.  In particular, \citet{Mutel1998}
presents results from 8.4 and 15 GHz observations of Algol obtained
using the Very Long Baseline Array (VLBA) in 1995.  The VLBA maps show
a double-lobed structure in the radio corona of the K star, which
appear to originate in or near the polar caps.  The quiescent radio
emission does not show signs of orbital modulation, again suggestive
of a corona concentrated on the poles of the K star.  Caution is
warranted in the interpretation of the spatial distributions of radio
and X-ray emission, however, since the former is produced by
gyrosynchrotron emission from relativistic electrons, and this
population might not be co-spatial with the thermal electrons
responsible for X-ray emission.  Direct X-ray observations present the
only means to probe the spatial distribution of this thermal
population.

In this paper, we primarily use the {\it Chandra} High Energy Transmission
Grating Spectrograph (HETGS) X-ray spectrum of Algol to test this
emerging picture of coronal activity.  We compare observed Doppler
shifts with the theoretical orbital velocities of the primary and
secondary stars of Algol, and compare the observed line widths with
theoretical expectations based on instrumental, thermal, and
rotational broadening.  Based on these comparisons, we are able to
place the first observational limits on the contribution of the B8 dwarf to
the X-ray emission of Algol, and to place constraints on the scale
height, and any non-thermal motions, characterizing the corona of
Algol~B.  Finally, we show that lines of O~VII seen from quadrature to
primary eclipse in a Low Energy Transmission
Grating Spectrograph (LETGS) observation of Algol support our conclusions
that the corona of Algol~B at temperatures of a few $10^6$~K must be
radially extended to at least one stellar radius.




\section{Observations}
\label{s:observations}

The {\it Chandra} HETGS employed for the observations used in this analysis 
consists of two gratings---the High Energy Grating (HEG) and
the Medium Energy Grating (MEG).  The HEG covers a wavelength range of
1.2-15 \AA\ (10.0-0.8 keV), with a typical line width of $\sim$0.012
\AA.  The MEG covers a wider wavelength range of 2.5-31 \AA\ (5.0-0.4
keV), with a typical line width of $\sim$0.023 \AA.  A more thorough
description of this instrumentation can be found in
\citet{HETGS1994}.  We made use of both HEG and MEG spectra.

The HETGS observation of Algol (ObsID 604) studied here was obtained
as part of the Guest Observer (GO) program and was undertaken in the
standard instrument configuration using the ACIS-S detector between UT
02:20 and 17:31 on 2000 April 1, for a total effective exposure time
of 54660~s.  In order to investigate the accuracy of the instrument
line response function, we also analyzed an observation of Capella
(ObsID 2583) obtained for routine calibration.  This observation was
also taken in the standard instrument configuration using the ACIS-S
detector between UT 17:46 and 02:03 on 2002 April 29 to April 30, for
a total exposure time of 29700~s.  We chose this particular
observation of Capella because it was obtained at an orbital phase of
$\sim$ 0.5, when the two giants were near conjunction and spectral
features were minimally broadened by orbital motion.

We also analyzed an LETGS observation of Algol (ObsID 2), in order to
examine the ratio of the O~VII forbidden to intercombination lines as
a function of orbital phase.  The data were observed  between UT 18:22
and 17:11 on 2000 March 12 to March 13, for 
a total exposure time of 82200~s.  
For details of the observation itself we refer to \citet{Ness2002}.  

\begin{figure} [p]
\epsscale{1}
\plotone{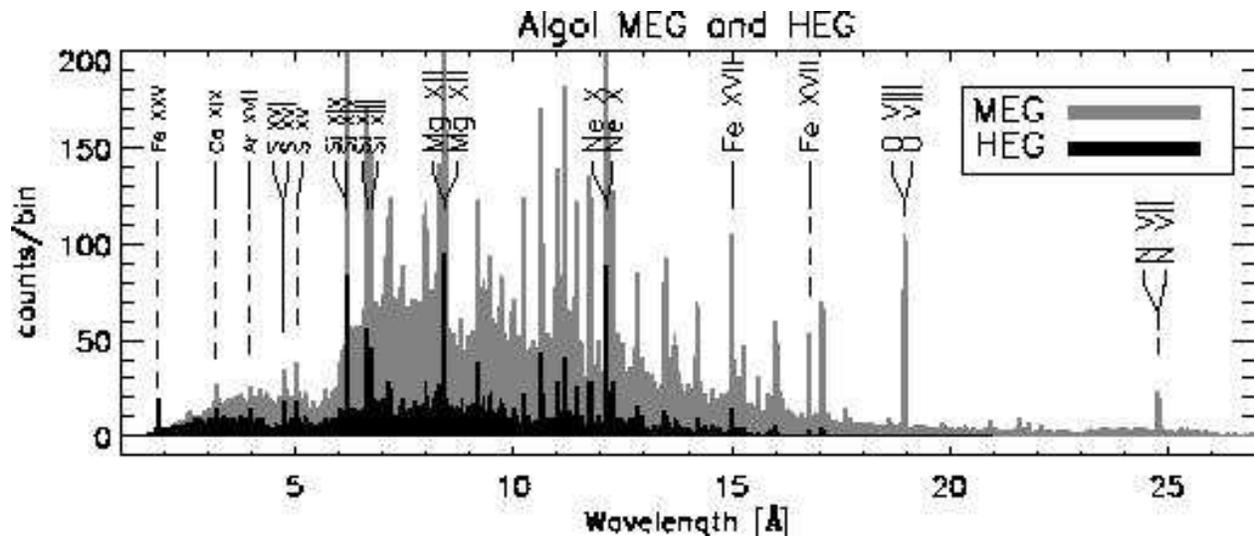}
\caption{This figure shows the coadded Algol spectrum, for both HEG
  and MEG.  Emission lines that were used in our analyses are labeled
  in a larger font size than lines that were not used.  The spectrum
  shown in the foreground is the coadded HEG spectrum, whose
  binsize is 0.0025 \AA.  The spectrum shown in the background is the
  coadded MEG spectrum, whose binsize is 0.005 \AA.}
\label{f:spec}
\end{figure}

The standard CIAO pipeline-processed (version 2.2) data were
downloaded from the Chandra public data archive.  Subsequent analysis
was done using the IDL software package PINTofALE
\citep{pintofale2000}.  The first order MEG and HEG spectra are
illustrated in Figure~\ref{f:spec}, together with identifications of
prominent spectral lines.  Emission lines used in our analyses are
labeled with a larger font.

\section{Analysis}
\label{s:analysis}

\subsection{Lightcurves}
\label{s:lightcurves}

Prior to spectral analysis, we examined the lightcurves of both Algol
and Capella in order to determine their level of variability and
whether any significant events (such as large flares) occurred during
the observations that may be relevant for subsequent interpretation of
Doppler shifts and line broadening.  The lightcurves were derived by
extracting all {\em dispersed} events in the standard CIAO spectral
extraction region, then binning the events at 100 second time
intervals.  We emphasize here that we did not use the 0th order events
since these are strongly affected by pile-up.

\begin{figure} [p]
\epsscale{1}
\plotone{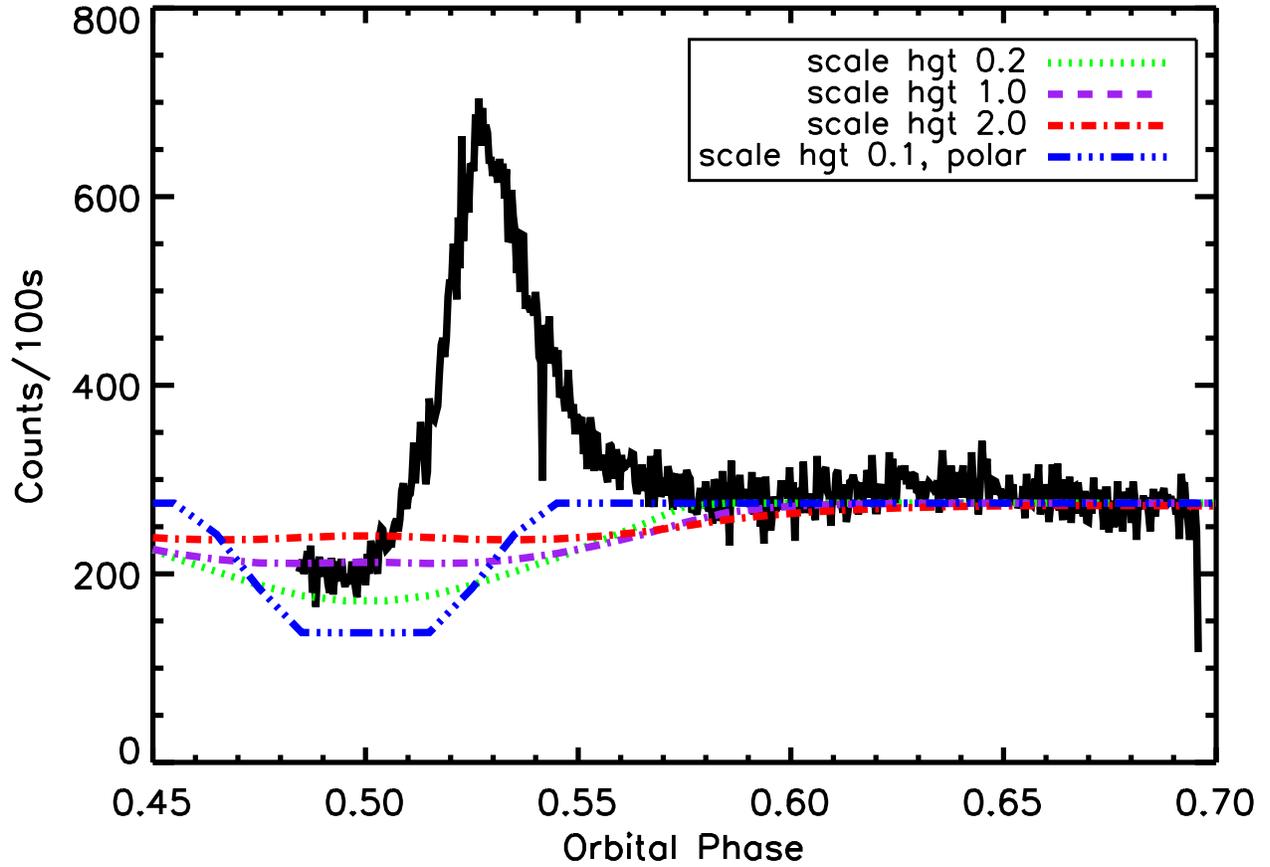}
\caption{The observed X-ray lightcurve of Algol derived from dispersed
  HEG and MEG events, with theoretical
  lightcurves overplotted.  The theoretical lightcurves are calculated
  for various K star coronal scale heights, and assume that Algol A is X-ray
  dark.  The observed lightcurve displays significant flare activity
  as Algol B comes out of eclipse.}
\label{f:lc}
\end{figure}

The observed lightcurve of Capella is relatively flat, with no obvious
signs of flaring, or other variability.  The lightcurve of Algol on
the other hand, shows a definite and significant flare in the
beginning of the 
observation (Figure~\ref{f:lc}).  Note that the orbital phase of the
Algol observation begins at $\phi$ = 0.48, just before Algol B starts
to come out of eclipse.  It is possible that the impulsive phase of
the flare has been affected by geometric occultation and we return to
this in \S\ref{s:results}. The flare has
decayed by $\phi=0.57$, soon after Algol B passes conjunction.  Algol
B appears to be quiescent for the remainder of the observation, which
ends at $\phi=0.70$.

\subsection{Doppler Shifts}
\label{s:doppler}

In this section, we investigate the Algol spectra for evidence of
Doppler shifts resulting from orbital motion.  
The following analysis is restricted to the emission lines listed in
Table~\ref{t:whichlines}.  We have chosen these particular lines 
because they have high S/N ratios and are not significantly blended
with lines of other atomic species.


\clearpage

\begin{table}  [h]
\begin{center}
\caption{A list of emission lines that were used for the analysis
  described in \S\ref{s:doppler}.  Rest wavelengths are taken from
  \citet{Chiantiversion4}.} 
\vspace{0.3in}
\small
\begin{tabular} {c c c c}
\hline
Wvl [\AA] & Ion & Grating & Diffraction Order \\
\hline
8.42 & Mg~XII & MEG & 1,3 \\ 
12.13 & Ne~X & MEG,HEG & 1 \\
15.01 & Fe~XVII & MEG & 1 \\
16.78 & Fe~XVII & MEG & 1 \\
18.98 & O~VIII & MEG & 1 \\
24.78 & N~VII & MEG & 1 \\ 
\hline
\label{t:whichlines}

\end{tabular}
\end{center}
\end{table}

\clearpage

Orbital velocity as a function of orbital phase is derived separately
for each emission line listed in Table~\ref{t:whichlines}.  For each
line, we first bin the events into time intervals (which are later
converted to orbital phase), where the bin size is proportional to the
total observed counts of the emission line, thereby maintaining an
approximately constant S/N ratio for each bin.  Negative and positive
order events (in both MEG and HEG) were combined to obtain higher
signal to noise (S/N) ratios.  Since we are trying to accurately
measure the wavelength centroids of emission lines, what is important
here is the S/N ratio, rather than the resolution.  Although the HEG
has better resolution than the MEG, most of even the brightest lines
proved too faint for useful analysis in HEG spectra.  Most of the
lines we analyze here are therefore MEG lines.

A modified Lorentzian function (`beta-profile') 
described by the relation
\begin{equation}
F(\lambda)=a/(1+(\frac{\lambda-\lambda_0}{\Gamma})^2)^\beta
\label{e:lorentz}
\end{equation}
where $a$
is the amplitude and $\Gamma$ is a characteristic line width, 
is then fit to the events along the
wavelength axis for each time interval.   
With $\beta=2.5$, this function has been found to be a good match to
observed HETGS line profiles, better than Gaussians 
\citet[][in preparation]{Drake2004inprep_a}. 
In this way, we
obtain the observed wavelength centroids as a function of time and
orbital phase.  The line-of-sight orbital velocities were derived for all
emission lines by subtracting the rest wavelengths of lines from their
observed centroids.  

We expect the dominant component of the Doppler shifts to result
from the orbital motion of Algol~B, which amounts to $\pm
200$~km~s$^{-1}$ at quadrature.   
The predicted orbital velocity is described by
\begin{equation}
v_{orb}=\frac{2\pi r \sin(i) \sin(2\pi\phi)}{P},
\label{e:vorb}
\end{equation}
where $r$ is the radius of orbit, $i$ is the inclination, $\phi$ is the
orbital phase, and $P$ is the orbital period.  
We used an inclination of 81.4 degrees \citep{Richards1993}
and a period of 2.87 days.  We fitted a sine
model to the observed line wavelengths using
the CIAO fitting engine {\tt sherpa}.  
The model used to fit the data is given by
\begin{equation}
f(\phi) = \delta \lambda_{o} +\frac{\lambda_{o}}{c}(A\sin[2\pi(\phi)]) 
\label{e:sinemodel}
\end{equation}
where $\delta \lambda_{o}$ is a constant y-offset [\AA], $\lambda_{o}$ is the
rest wavelength of the emission lines [\AA], $c$ is the speed of light
[km~s$^{-1}$], $A$ is the amplitude of the final fit which represents the 
line-of-sight orbital velocity [km~s$^{-1}$] (given by 2$\pi$r~$\sin i$/P),
$\phi$ is the orbital phase, r is the effective orbital radius of
X-ray emitting material [km], $i$ is the inclination of orbit [radians], 
and P is the orbital period [s].
The parameters for all emission lines were forced to be the same, and 
the orbital period of Algol~B was fixed at the value noted above.  The
only parameters allowed to vary were $\delta \lambda_{o}$ and
$A$.  
We allowed $\delta \lambda_{o}$ to be a free parameter because the
absolute wavelength calibration of HETG+ACIS-S observations is
$\sim$~0.011 \AA\ for MEG and $\sim$~0.006 \AA\ for
HEG\footnote{http://asc.harvard.edu/proposer/POG/html/HETG.html}.
Therefore, each emission line may need a different  $\delta
\lambda_{o}$  offset
in order to account for wavelength calibration
uncertainties.   Because $\delta \lambda_{o}$ was 
a free parameter, errors on the reference wavelengths, though
extremely small for the H- and He-like ions, 
are not relevant for our analysis.  
The data for all spectral lines were fit simultaneously. 

\begin{figure} [p]
\epsscale{0.9}
\plotone{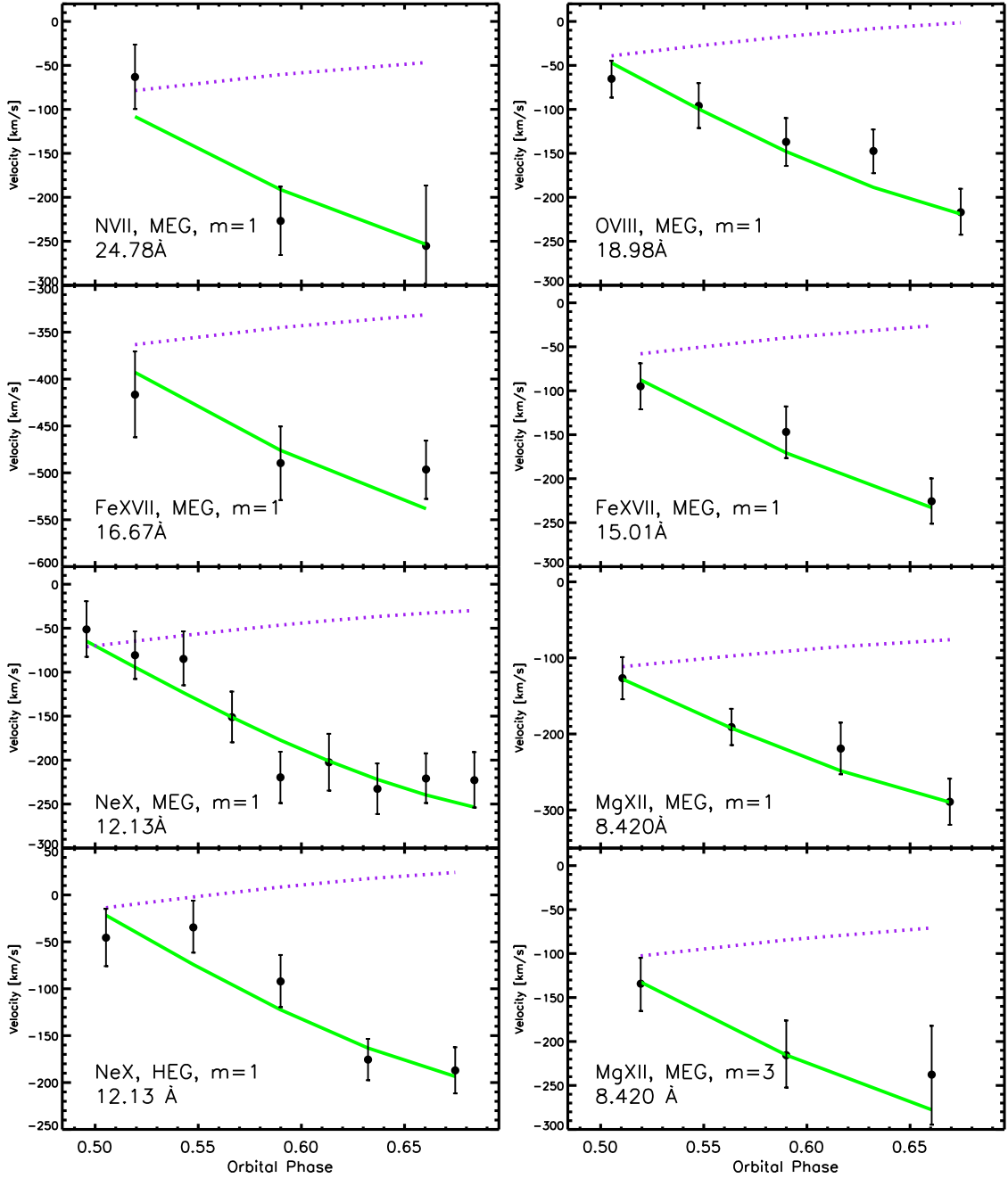}
\caption{This plot shows the observed 
  line-of-sight orbital velocity and 1$\sigma$ error bars as a function
  of orbital phase for all lines listed in Table 1.
  Overplotted are the theoretical orbital velocities of Algol~A
  (dotted line) and Algol~B (solid line).
  The theoretical velocities have been vertically offset to best fit
  the data, according to the $\delta \lambda_{o}$ obtained from the
  {\it sherpa} fitting (equation~\ref{e:sinemodel}).}
\label{f:orbit_all}
\end{figure}

The amplitude of the best fit was A=187.5~km~s~$^{-1}$.  From the
amplitude, we can compute the effective orbital radius of X-ray
emitting material, $r$ (equation~\ref{e:vorb}).  We obtain an
effective orbital radius of 10.74 $\pm$ 0.93 R$_{\odot}$.  This result
will be further discussed in \S\ref{s:results}.
Figure~\ref{f:orbit_all} illustrates the observed line of sight
orbital velocity as a function of orbital phase, derived from various
emission lines.  Figure~\ref{f:orbit_all} also shows the theoretical
orbital velocities of Algol A and Algol B, calculated under the
assumption that the X-ray emission is centered on one of the two stars
(i.e.--the expected curve if {\em all} of the X-ray emission had been
located at the poles of Algol B or Algol A).

\subsection{Cross-Correlation: Doppler Shifts}
\label{s:cross-cor}

In addition to measuring individual emission lines, we also utilized a
cross-correlation technique to obtain Doppler shifts as a function of
orbital phase.

Events were binned into 5 time intervals corresponding to different
orbital phases, and spectra were
extracted for each phase bin.
Spectra from each phase bin were cross-correlated with a reference
spectrum whose orbital velocity is known.  We used Capella (ObsID
2583) as our reference spectrum because, as noted in
\S\ref{s:observations}, this particular observation
was taken when the two giants were near conjunction in orbital phase.
This ensures that the orbital velocity of the spectrum is nearly zero.
Other systematic velocities involved can all be determined with high
precision and subtracted: these are the radial velocities of the
Capella and Algol systems, and the velocity of the {\it Chandra}
satellite in the direction of these objects during the observations.
Algol has a radial velocity of only 4~km~s$^{-1}$ \citep{Wilson1953},
while Capella has a radial velocity of 30~km~s$^{-1}$
\citep{Wilson1953}, therefore we must redshift the Algol velocities by
26~km~s$^{-1}$.  The orbital velocity of {\it Chandra} during these
observations was less than $\sim$2~km~s$^{-1}$.  Since this velocity
is small compared to the errors incurred during the cross-correlation
analysis, we ignored this orbital motion.



The Algol and Capella spectra were cross-correlated by shifting the
Algol spectrum with respect to the Capella spectrum in velocity steps
of 25~km~s$^{-1}$, ranging from -200 to 500~km~s$^{-1}$.  At each
velocity shift, we computed the $\chi^{2}$ value as follows:
\begin{equation}
\chi^{2} = \sum
\frac{[C_{ref}(\lambda)-C_2(\lambda(1+\frac{v}{c}))]^2}{\sigma_{ref}(\lambda)^{2}+\sigma_{2}(\lambda(1+\frac{v}{c}))^2};
\label{e:chi2}
\end{equation}
where $C_{ref}(\lambda)$ and
$C_2(\lambda(1+\frac{v}{c})^2)$ are the two spectra being
compared (Capella and velocity shifted Algol), and
$\sigma_{ref}(\lambda)$ and
$\sigma_{2}(\lambda(1+\frac{v}{c})^2)$ are their
respective errors.  These quantities are then summed over the entire
spectral region.  The velocity shift at which minimum $\chi^{2}$ is
achieved represents the Doppler shift in that particular phase bin of
the Algol spectrum relative to the Capella spectrum.

In order to better understand the errors incurred from
cross-correlating the two spectra and computing $\chi^{2}$, we
utilized a Monte Carlo method by randomizing the counts within errors
on the Algol spectrum 25 times, and repeating cross-correlation for each
randomized data set. Parabolae were then fit to $\chi^{2}$ versus
velocity shift  for each of the 25 simulations in order to 
find the velocity that corresponds to the minimum $\chi^{2}$
of each simulation.  The mean velocity is then adopted as the final
Doppler shift for that given interval of phase, and the standard
deviation of the mean is the $1 \sigma$ error on the final Doppler
shift.  

\begin{figure} [p]
\epsscale{1}
\plotone{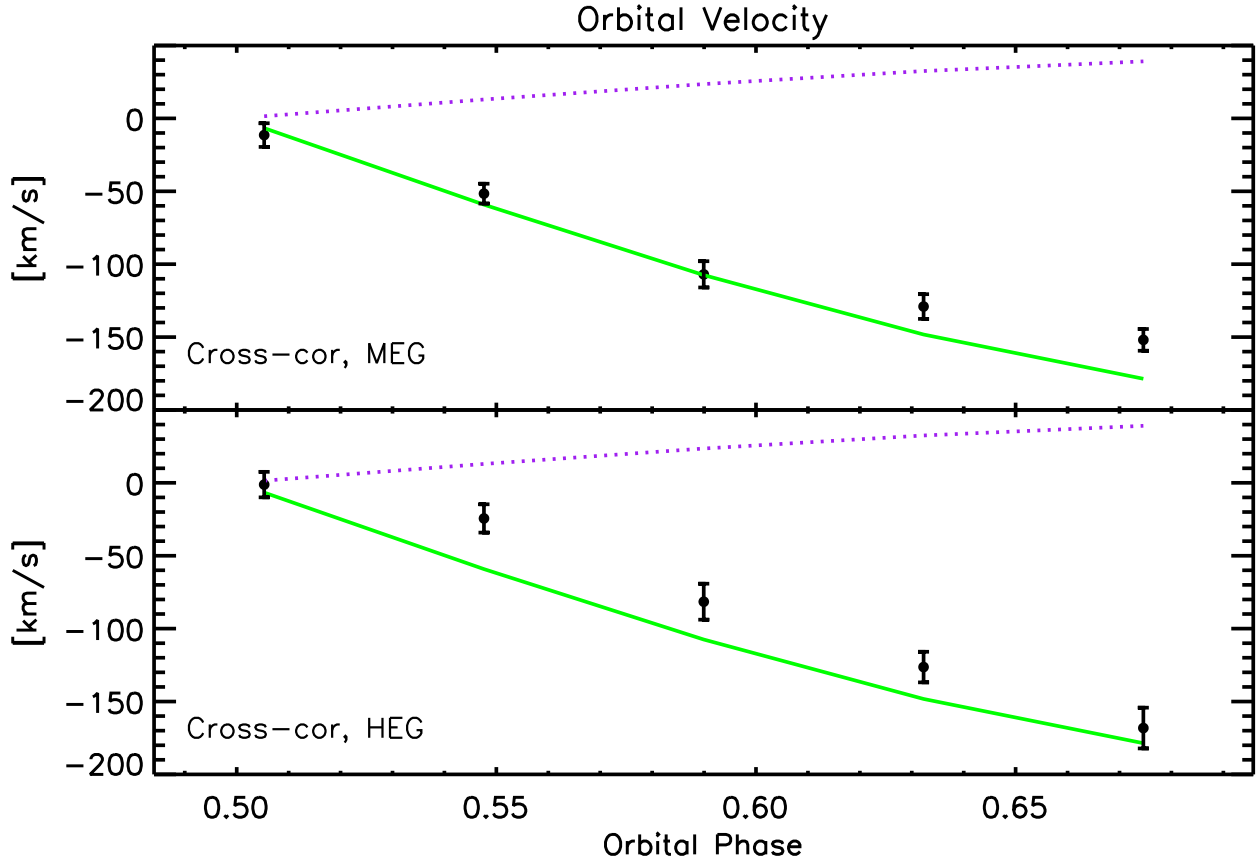}
\caption{Illustrates the line of sight orbital velocity as a function of
  orbital phase for MEG and HEG data, obtained by cross-correlating
  the Algol spectrum
  with respect to Capella.  We also show the theoretical orbital velocities of
  the Algol primary (dotted line) and secondary (solid line) stars.}  
\label{f:orbit_crosscor}
\end{figure}

Figure~\ref{f:orbit_crosscor} illustrates the line-of-sight
velocity of Algol as a function of orbital phase, as obtained by the
cross-correlation/monte-carlo analysis method described above.  Also
illustrated are the theoretical orbital velocities of Algol A and
Algol B.



Once again, we used the {\tt Sherpa} fitting engine to fit a sine
model to the observed Doppler shifts obtained via the
cross-correlation analysis.  Doppler shifts for the positive and
negative orders, as well as both MEG and HEG, were fit simultaneously
with the model described by equation~\ref{e:sinemodel}.  The amplitude
of the sine model is A=173.4 $\pm$ 5.1~km~s$^{-1}$.  Assuming an
orbital period of 2.87 days, and an inclination angle of 81.4 degrees,
this amplitude corresponds to an orbital radius of 9.93 $\pm$ 
0.29$R_{\odot}$.  This radius is consistent, within 1$\sigma$
uncertainties, with our previous result of 10.74 $\pm$ 0.93$R_{\odot}$ 
obtained by measuring wavelength centroids of
individual emission lines.

We also applied the
cross-correlation/monte-carlo analysis to the Algol spectrum for
regions greater than and less than 10 \AA, in order to compare the
line-of-sight velocities for lines formed at cooler versus hotter
temperatures.  Doppler shifts for both regions of the spectrum were
consistent, though uncertainties for the hotter region of the
spectrum were considerably larger ($\sim$50~km~s$^{-1}$) than the
cooler region ($\sim$10~km~s$^{-1}$), since most of the stronger
emission lines lie at wavelengths greater than 10 \AA.  

\subsection{Thermal and Non-thermal Broadening}
\label{s:thermal}

\subsubsection{Testing the Instrumental Profile using Capella}
\label{s:capella}


In order to test for the presence of non-thermal broadening in the
observed Algol line profiles, we compare these with carefully-computed
theoretical profiles.  Since the dominant source of line-broadening in
the observed spectra is the instrumental profile, it is important that
we confirm that our understanding of the instrumental broadening is
correct, before
proceeding with the comparison.

In a far ultra-violet study of Capella,
\citet{Young2001} found that the observed profile of Fe~XVIII
originated largely from the G8 component, and that only instrumental,
rotational (3~km~s$^{-1}$), and thermal broadening were required to
match the observed profile.  This gives us some confidence that the
coronal spectra of the Capella giants are free from significant
additional sources of non-thermal broadening.  It is not certain which
of the two Capella stars dominated during the time of the observation
analyzed here; for example, 
\citet{Johnson2002}
found that, in 1999 September 12, the G1 component dominated in the
light of Fe~XXI~1354 \AA\ seen by the Space Telescope Imaging
Spectrograph (STIS).  The dominance of one component over the other is
not important here.  By constructing two theoretical profiles
(described below), one with rotational broadening of 3~km~s$^{-1}$ (G8
component) and another with rotational broadening of 36~km~s$^{-1}$
(G1 component; \citep{Griffin1986}), we find that the difference in
FWHM of the two
profiles are negligible---of order 0.0001 \AA\ and two orders of
magnitude smaller than typical observed line widths.


Theoretical line profiles were synthesized by convolving the
predicted rotational, thermal, and instrumental line broadening, where
rotational profiles were derived by assuming solid body rotation at
the stellar surface \citep[e.g.][]{Gray1992}.  In our final synthesis
of Capella theoretical line profiles, 
we have adopted a rotational velocity value of the
G1 star (36~km~s$^{-1}$; \citep{Griffin1986}).  
The width of the thermal profile is calculated as the weighted average
of the thermal widths computed at various temperatures where the
weighting function is given by the product of the emission measure and
the line emissivity at each temperature.  The FWHM of the thermal
width for a given ion, at a given temperature, is then described by:
\begin{equation}
v_{thermal} = 2.335 \times (\Phi(T)\times \epsilon(T)\sqrt{\frac{2kT}{m}})
\label{e:thermal}
\end{equation}
where $\Phi(T)$ is the emission measure,
$\epsilon(T)$ is the emissivity, $T$ is the electron 
temperature, $k$ is the
Boltzmann constant, and $m$ is the ion mass.
We derived an emission measure $\Phi(T)$ for Capella based on the
measured line strength ratios of H-like and He-like ions, as described
by \citet[][in preparation]{Drake2004inprep}.  
This method of using line strength ratios to derive an emission
measure is similar to what is described in
\citet[][in press]{Schmitt2004_EM}. 
We defer a detailed discussion of this analysis to a future
paper, but emphasize that the final thermal profile does not depend
critically on the exact details of the emission measure distribution:
choosing a single isothermal emission measure at a temperature of
$\log T=6.8$ produced very similar results.

Line profiles for the H-like Lyman-$\alpha$ doublets of O~VIII, Ne~X, 
and Mg~XII were each modeled with the doublet 
wavelength separation fixed at $\Delta \lambda_{doublet} =
0.0054$ for Ne~X and Mg~XII, and $\Delta \lambda_{doublet} = 0.0055$ for
O~VIII \citep{Chiantiversion4}.  Both components of a given line
were forced to have the same line width, and relative fluxes of the
two components were fixed at the theoretical ratio of 2:1.

\begin{figure} [p]
\epsscale{1}
\plotone{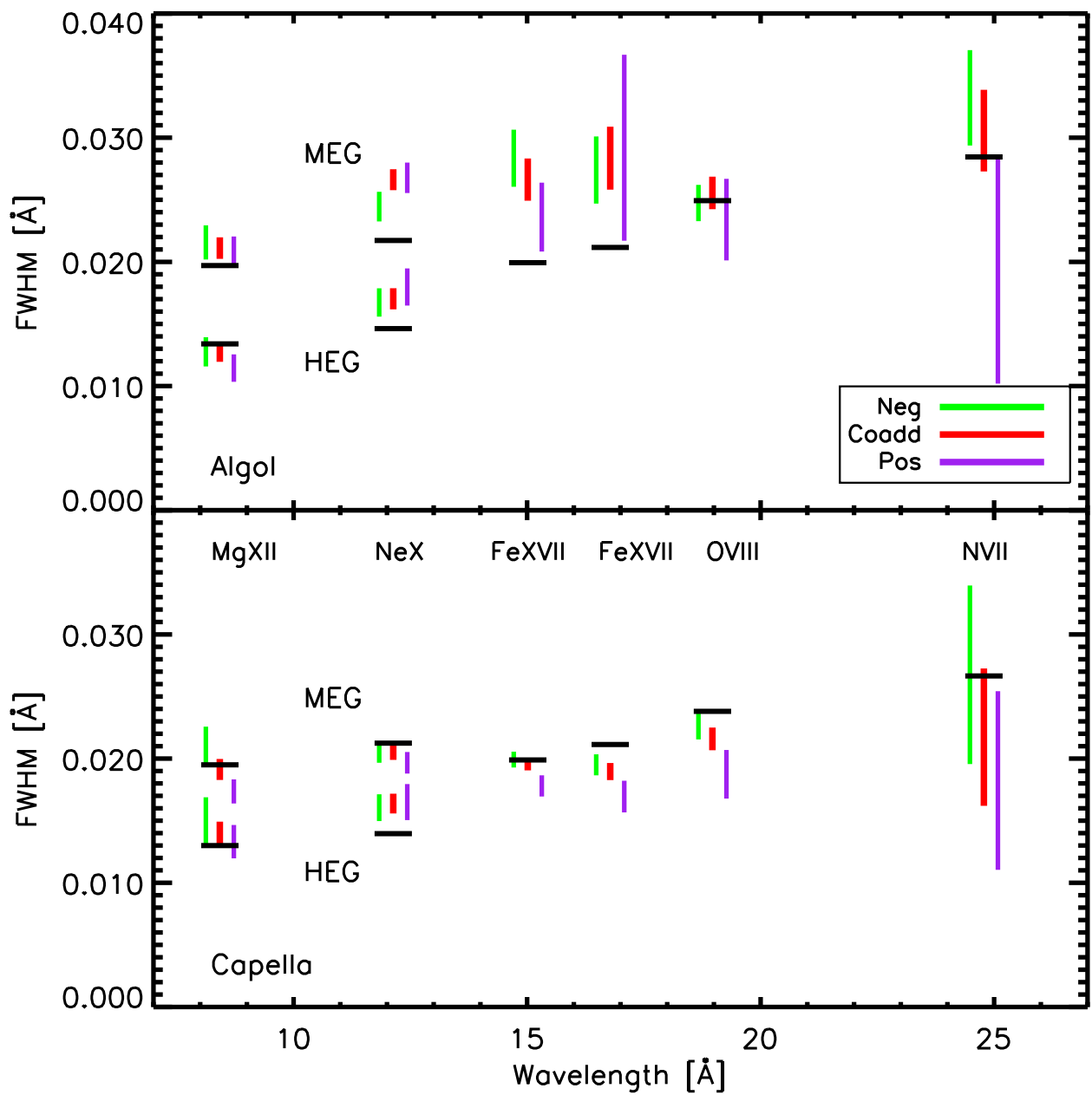}
\caption{Illustration of the observed line widths of six emission
  lines for negative, positive, and coadded orders, for Algol (upper
  panel) and Capella (lower panel).  The negative and positive order
  wavelengths were shifted to the left and right respectively, 
  so as to be more clearly
  visible.  The solid horizontal line indicates predicted line
  widths.  While the  theoretical and observed line widths
  are consistent with each other for Capella,
  the Algol observed line widths show a
  significant excess as compared to theoretical values.}

\label{f:order_comp}
\end{figure}

The final theoretical line profiles are then created by convolving the
rotational and thermal profiles with the expected instrumental
broadening.  The instrumental line profile was derived from
high-fidelity ray-traces of the {\it Chandra} instruments \citep[][in
press]{Marshall2004}.  The predicted and observed line widths are
compared in Table~\ref{t:linewidths} and Figure~\ref{f:order_comp}.
These comparisons show that the theoretical Capella line widths are in
good agreement with the observed line widths of the data.  We are
therefore confident that the current model of the instrumental line
broadening is sufficient to yield theoretical profiles that are
accurate to within errors of the observed Algol line widths.

\subsubsection{Algol Line Widths}

Prior to measuring the Algol line widths, we eliminated line
broadening due to the orbital velocity of Algol B by shifting the
event positions by the predicted amount ($v = 2\pi r\sin i/P$).  Here,
we adopt the true radius of orbit of Algol~B, $r = 11.5R_{\odot}$; we
also performed the same line width analysis assuming the smaller of
the two effective radii, $r=9.93 R_\odot$ found from the Doppler
shifts above, but found no significant differences between the two.
The negative and positive order spectra were then coadded.  Great care
has to be taken here.  Bright sources such as Algol suffer from
significant pile-up in the 0th order, resulting in a distorted source
profile.  This distortion can confuse the standard pipeline 0th order
location algorithm, resulting in a computed centroid that does not
correspond exactly to the true centroid.  If the 0th order centroid
location is not precise, there will be a systematic wavelength shift
between opposite orders.  Co-addition of the opposite orders will then
produce an artificially broadened line profile.  In order to avoid
this problem, we therefore ensured that the line centroids in opposite
spectral orders were identical prior to co-addition, on a line by line
basis.  The widths of lines listed in Table~\ref{t:whichlines} were
then measured by fitting with a modified Lorentzian function
(equation~\ref{e:lorentz}).

Since the measured line widths are sensitive to the adopted level of
the continuum, we have computed a model continuum based on the derived
emission measure distribution of Algol and have used this to guide
continuum placement.  In this way, the continuum level adopted makes
use of the signal in the spectrum over a broad wavelength range and
accurate placement can be achieved such that final uncertainties in
measured line quantities do not have a significant contribution from
continuum placement errors.  The final continuum placement was done by
eye, using effectively ``line free'' regions as a guide.  In order to
verify that this continuum placement did not introduce significant
errors in our line parameters, we performed a sensitivity test by
changing the adopted continuum level by different amounts that
stretched plausibility regarding the true continuum level, and
re-measuring lines for the different levels.  Although the continuum
placement is partly subjective because of the pseudo-continuum due to
the presence of weak lines and broad wings of closely spaced lines, in
the worst case the uncertainty of the FWHM of any of our lines
resulting from uncertainties in continuum placement is not more than
$\sim$0.002 \AA .  For some of the stronger lines, such as Ne~X, it is
unlikely that the line width uncertainty is at most $\sim$0.001 \AA .


Theoretical profiles for Algol were generated in the same way as
described for Capella in \S\ref{s:capella}.  
The theoretical FWHM were then compared to the
measured line widths of the observed data; these comparisons are
listed in Table~\ref{t:linewidths} and illustrated in
Figure~\ref{f:order_comp}.  We find some evidence for moderate excess line
broadening in several of the emission lines.  In the MEG spectra, all
lines but N~VII, O~VIII, and Mg~XII show signs of significant excess line
widths.  The two Fe~XVII lines, Ne~X, and Mg~XII show significant
excess line widths, ranging from 3$\sigma$ to almost 6$\sigma$.  In
the HEG data, Ne~X shows an excess width of 3$\sigma$, though
Mg~XII shows no sign of excess line broadening.  

In order to determine whether any of the observed excess broadening
could be due to the flare activity in the first part of the
observation we also performed the analysis described above for flare
and quiescent periods.  Line widths for both periods were found to be
consistent within statistical uncertainties.


\subsubsection{Turbulent or Flow Velocities}
\label{s:turbvel}

The excess widths we have detected in Algol may be attributed to
either or both of two plausible sources of non-thermal velocities: 
turbulence or flows within coronal structures, and ``excess''
rotational broadening above what is expected from surface emission.  

\begin{figure} [p]
\epsscale{1}
\plotone{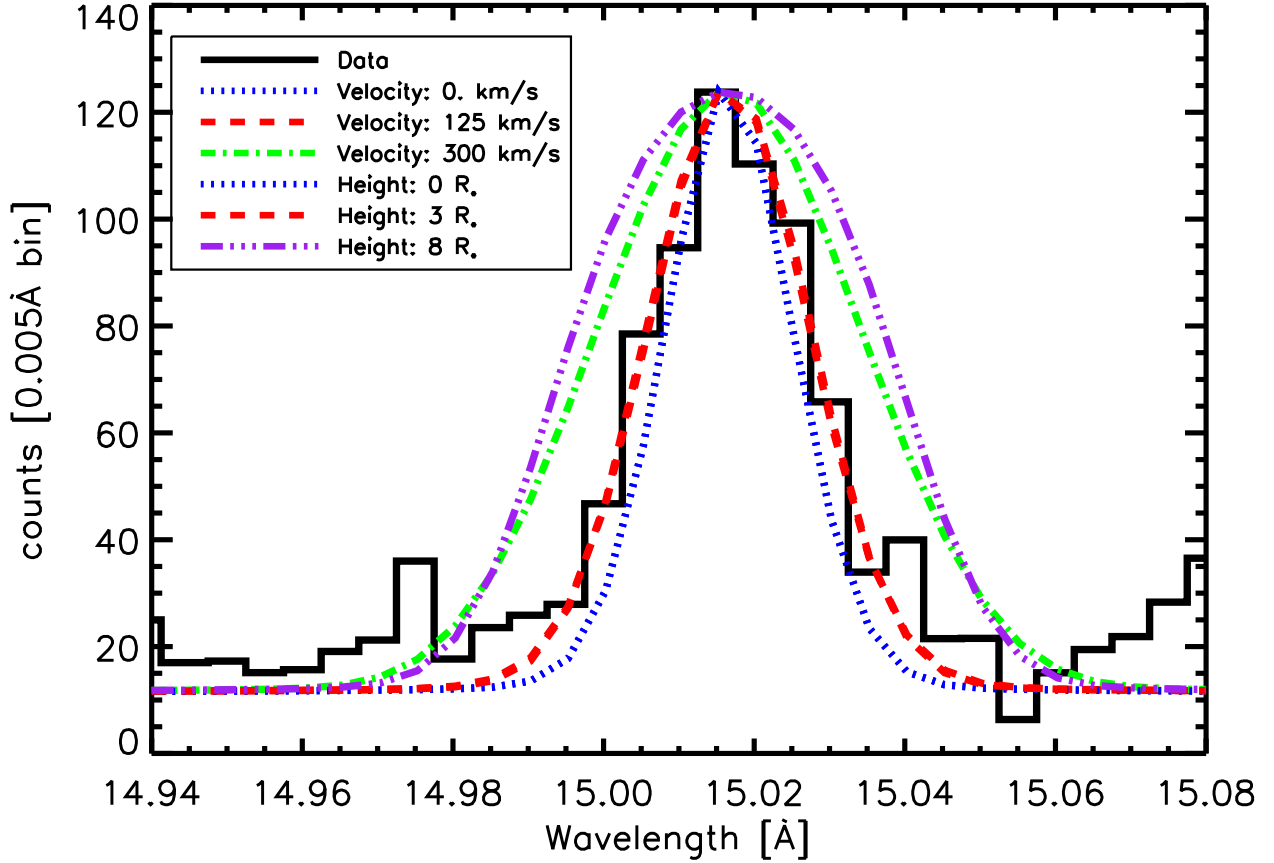}
\caption{
The observed profile of the prominent Fe~XVII
15.01~\AA\ resonance line overplotted with various theoretical
profiles.  The theoretical profiles were all normalized to the peak of
the observed data.  They are not actual fits to the data, but serve to
illustrate what approximate velocities and scale heights best match
the data.  ``Heights'' are expressed relative to the stellar surface
(i.e., a scale height of 0 represents rotational broadening at the
stellar surface).
Profiles computed for different velocities assumed a height of
$0~R_\star$; profiles computed for different scale heights assumed a
non-thermal velocity of 0~km~s$^{-1}$.  Note that the profiles for a
velocity of 125~km~s$^{-1}$ and a scale height of $4R_\star$ are
essentially identical and are represented by the same profile.}
\label{f:profile}
\end{figure}

Figure~\ref{f:profile} is an example comparison between observed and
theoretical line profiles.  This figure shows that a theoretical
profile with an additional Gaussian velocity of 125~km~s$^{-1}$ is a
reasonably good match to the Fe~XVII line, while an additional velocity of
300~km~s$^{-1}$ is too wide, and a velocity of 0~km~s$^{-1}$ is too narrow.

\begin{figure} [p]
\epsscale{0.9}
\plotone{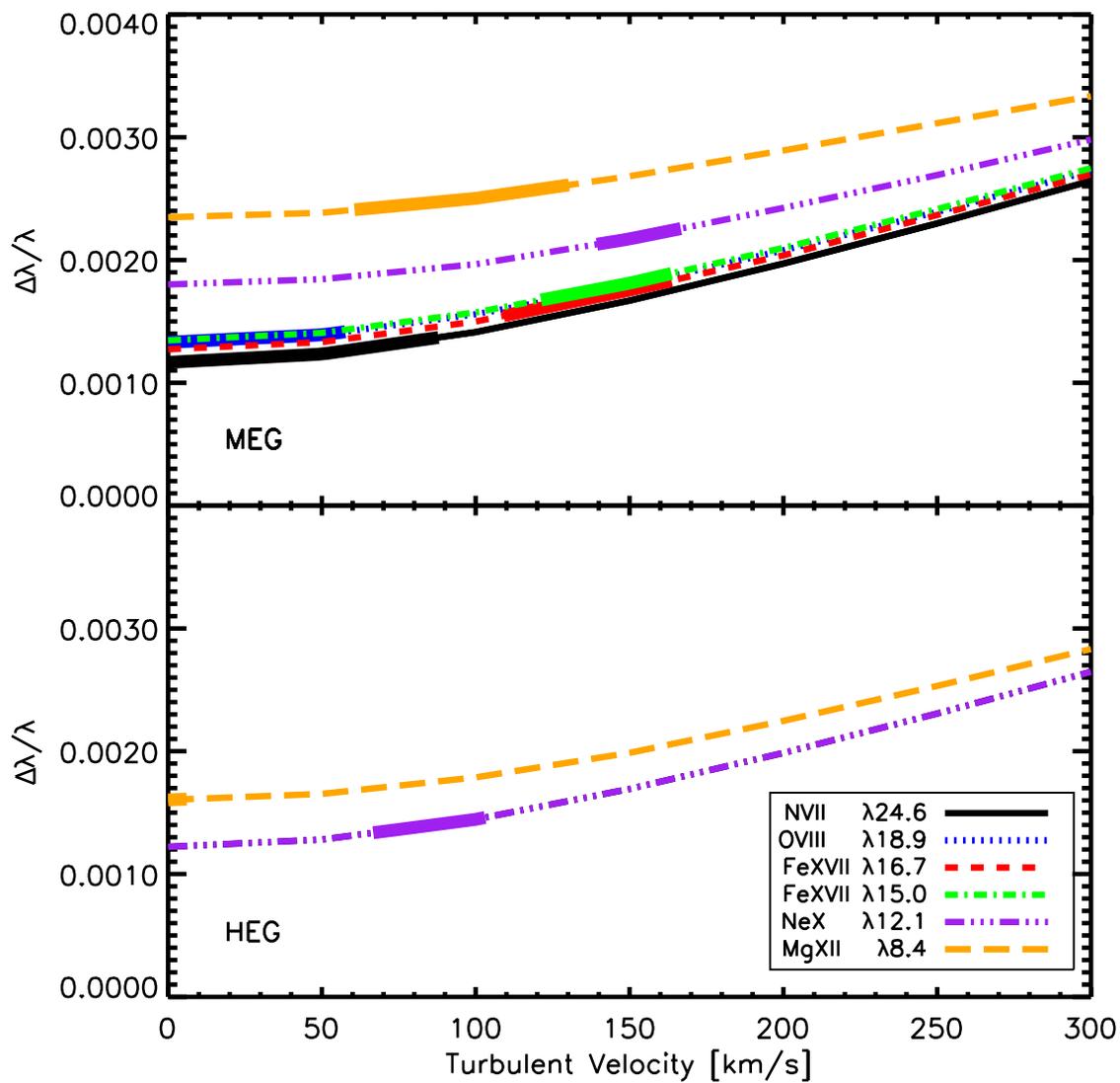}
\caption{Theoretical FWHM divided by rest wavelength as a function of
  additional Gaussian velocity.  Overplotted in thick lines are the
  observed Algol line widths divided by the rest wavelengths, with
  $\pm$~1$\sigma$.  
  The upper panel shows MEG data,
  and the lower panel shows HEG data.}  

\label{f:turbulent}
\end{figure}

We obtain a crude estimate of the turbulent velocities
involved by comparison of observed and theoretical line profiles,
computed as described in the previous section.  We convolved the
latter with an additional Gaussian broadening corresponding to a range
of velocities from 0 to 300~km~s$^{-1}$.  Figure~\ref{f:turbulent}
illustrates the FWHM of these synthesized profiles as a function of
the additional non-thermal velocity, and also indicates where the
observed FWHM fall on these curves.  

Most of the lines we have analyzed show observed FWHM that correspond
to additional velocities of $\sim$ 50-150~km~s$^{-1}$. 
We are able to place an upper-limit on additional turbulent velocities
at \lap 170~km~s$^{-1}$.

\subsubsection{Rotational Broadening}
\label{s:rotational}

Excess rotational broadening can occur if the corona on
Algol B is significantly radially extended.  Line widths can
therefore be used to place direct spectroscopic constraints on this
radial extension.  Limits on the radial extent of the corona of
Algol~B were determined by comparison of observed and theoretical
profiles computed for different scale heights.  We varied the scale
height of the theoretical rotational profiles from within the stellar
surface (zero rotation) to 8.75$R_{B}$, where the scale height is
defined as zero at the
stellar surface.  Then we convolved these with the thermal and instrumental
broadening, as before. The final profiles are compared with the
Fe~XVII~15.01~\AA\ resonance line in Figure~\ref{f:profile}.  This
figure shows good qualitative agreement with a profile computed for
coronal scale height of 3$R_{B}$; a scale height of 8$R_{B}$
is clearly too broad, while purely surface rotational broadening is
too narrow.

\begin{figure} [p]
\epsscale{0.9}
\plotone{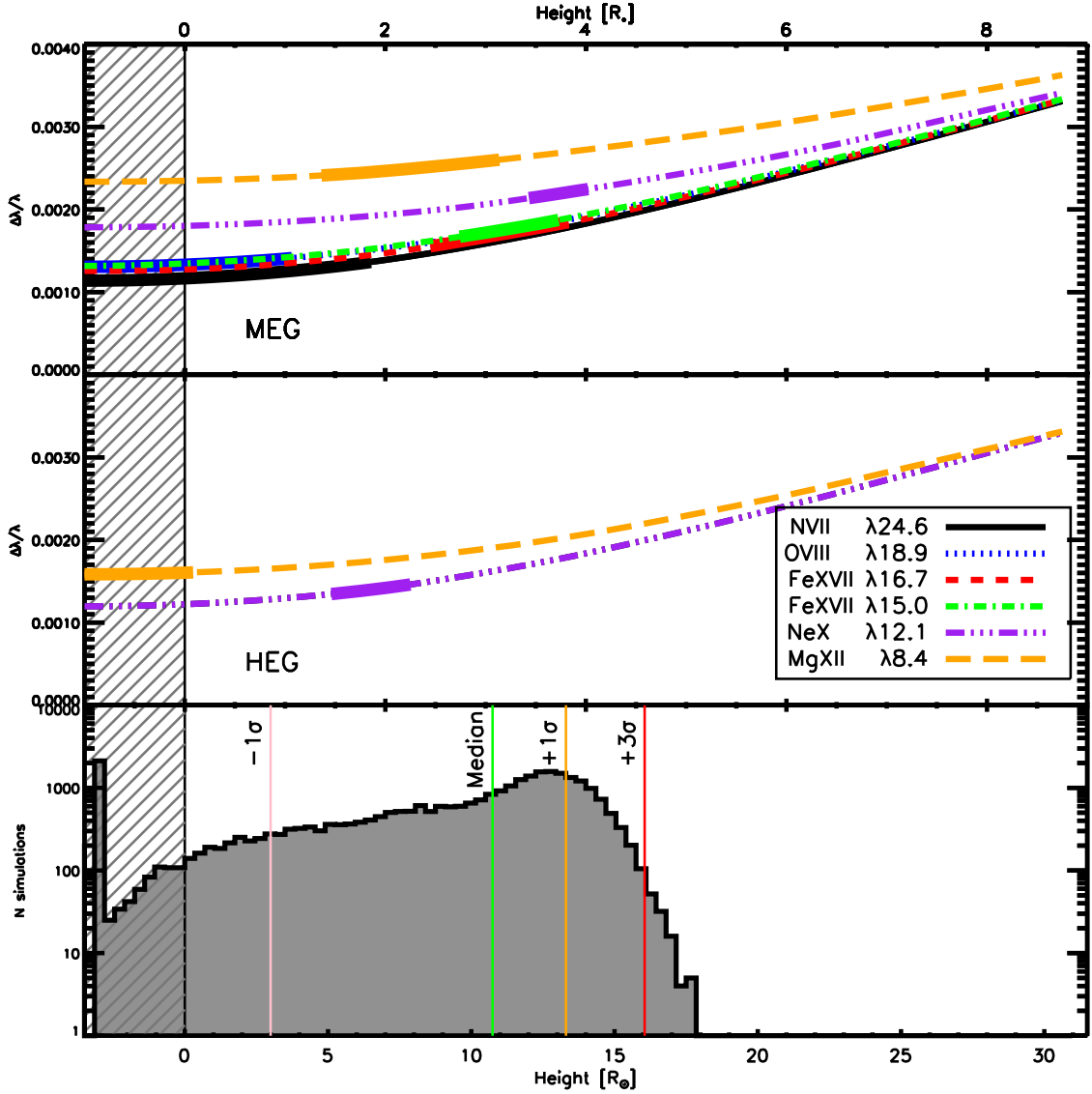}
\caption{The top 2 panels show the 
  theoretical FWHM divided by rest wavelength as a function of 
  Algol B coronal scale height, for MEG and HEG data, respectively.  
  Lower x-axis indicates Algol B scale
  height in units of solar radii, while upper x-axis indicates scale
  height in units of stellar radii.  The darkened region of the plot
  encloses the area where the rotational broadening is less than or
  equal to the rotational broadening at the surface of Algol B.
  Overplotted in thick lines are the Algol observed line widths
  divided by the appropriate rest wavelength, with $\pm$~1$\sigma$
  error bars on each side.  
  The bottom panel shows the combined data from all lines, as realized
  by a monte carlo sampling of the scale heights and uncertainties
  indicated by the different spectral lines in the top two panels. 
  theoretical curves shown in the top two panels.  The vertical
  lines indicate the median, the 1$\sigma$ and 3$\sigma$ upper
  limits, and the 1$\sigma$ lower limit.}
\label{f:scale}
\end{figure}

The upper two panels of Figure~\ref{f:scale} show the FWHM of the 
theoretical profiles as a
function of coronal scale height, together with the observed FWHM
overplotted in thicker lines.  More than half of the observed FWHM fall
approximately at a scale height of $\sim$~2 to 3$R_{B}$, which
corresponds to an excess rotational velocity of
$\sim$~125 to 185~km~s$^{-1}$, in addition to the rotational 
velocity at the stellar surface ($\sim$~62~km~s$^{-1}$).

The bottom panel of Figure~\ref{f:scale} shows a distribution of
scale heights which were derived by randomly generating a set of line
widths that are distributed in a Gaussian manner, centered on the
measured line width of each emission line.
The number of monte-carlo realizations for each emission line was 
weighted by the
measured flux of that line.  We generated a total of 30087
realizations for all lines combined.
We then obtained the scale height that corresponds to each realization
of a line width by interpolating along the theoretical curves 
shown in the
top two panels of Figure~\ref{f:scale}.  The lowest panel of
Figure~\ref{f:scale} shows the distribution of the scale heights for
all realizations of line widths, for both MEG and HEG.
Vertical lines indicate the median scale height, and the 
1$\sigma$ and 3$\sigma$ limits.  The median indicates a scale
height of 3.1$R_{B}$, with 
1$\sigma$ and 3$\sigma$ upper limits occurring at scale heights of 3.8 and
4.6 stellar radii, respectively.  The lower 1$\sigma$ limit occurs at
0.9$R_{B}$. 
In the case of line width realizations
that are smaller than what is physically possible, these 
scale heights have been set equal to -1$R_{B}$, which accounts for the
large peak at -1$R_{B}$.

\subsubsection{O~VII $f/i$ ratio}
\label{s:f/i_analysis}

The UV excitation of the $2\,^3P$ term by Algol~A provides a possible
test of coronal geometric extent.  In the spherically-symmetric case,
a very compact corona should exhibit modulation of the ratio of the
forbidden and intercombination lines, $f/i$, with orbital phase: at
phase $\phi=0$---primary optical eclipse---the visible hemisphere of
Algol~B is not illuminated by Algol~A and the $f/i$ ratio should
revert very closely to its pressure-dominated value.  The active star
survey of \citet[][in preparation]{Testa2004} 
indicates that this ratio should be
$f/i\sim 2$-3, instead of 0.94, as observed by \citet{Ness2002}.

There is no evidence for orbital modulation of the $f/i$ ratio in the
HETGS observation; this is not surprising because the observation
covered only those phases during which the visible K star corona was
substantially illuminated by Algol~A.  We have therefore re-examined
the LETGS observation (ObsID 2) that covers phases $\phi=0.75$-1.03 to search
for evidence of orbital-modulated changes in the $f/i$ ratio.  The end
of this observation corresponds to primary eclipse when the visible
hemisphere of the K star is not irradiated by Algol~A.

\begin{figure} [p]
\epsscale{1}
\plotone{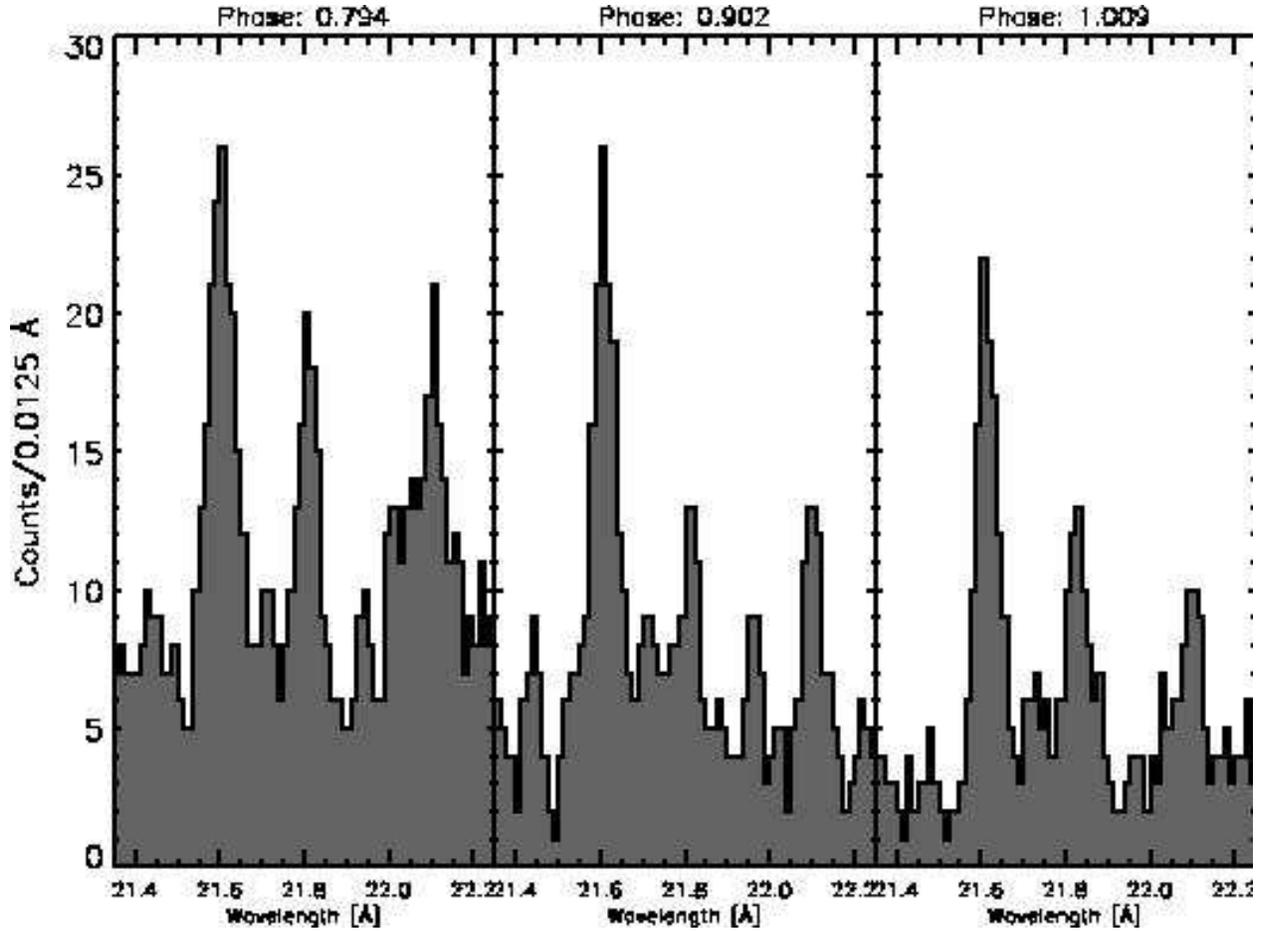}
\caption{O~VII lines extracted from three orbital phase intervals of an
  LETGS observation of Algol.  The intensity ratio of the
  forbidden to intercombination line ($f/i$) appears constant as
  Algol B moves from quadrature to conjunction, eclipsing the primary
  star at phase~$\sim$~1.}
\label{f:o7letg}
\end{figure}

We split the Algol LETGS data into three phase bins (lacking
the signal in the O~VII lines to make more fine divisions) and
examined the relative strengths of the intercombination and forbidden
lines at 21.80~\AA\ and 22.10~\AA .  The spectral region containing
these lines is illustrated for the three phase bins in
Figure~\ref{f:o7letg}:  there is no perceptible change in the ratio
between these different bins.  

\section{Discussion}
\label{s:results}

\subsection{Orbital Velocity}

By modeling observed Doppler shifts in terms of orbital motion,
we were able to estimate the effective orbital radius of the
X-ray emitting material.  We found this effective radius to be
$R_{eff}=10.74$~$\pm$~0.93$R_{\odot}$ from the analysis of individual
emission lines, and $R_{eff}=9.93$~$\pm$~0.29$R_{\odot}$ from the
cross-correlation analysis; these results are consistent within
$1\sigma$ uncertainties.  The statistical uncertainties in the 
cross-correlation
method are considerably lower because all the information in the
spectrum is used, rather than just information in bright lines.
The 1$\sigma$ error bar of the line-of-sight velocities obtained
via the cross-correlation analysis is
$\sim$8~km~s$^{-1}$ for the MEG data and $\sim$11~km~s$^{-1}$ for the 
HEG data.  
However, limitations due to uncertainties in the calibration of the
{\it Chandra}
instruments cause systematic uncertainties which appear somewhat 
larger than what
is implied by the statistical uncertainties.  For example, a few of
the data points in Figure~\ref{f:orbit_crosscor} show discrepancies
between the MEG and HEG results, by $\sim$2$\sigma$.  

In comparison with the X-ray Doppler radius, the orbital radius of
Algol B around the system center of mass is $R_{orbit(B)} = 11.5R_{\odot}$
\citep{Richards1993}.  Since the effective radius we have derived 
is smaller than $R_{orbit(B)}$, we can infer that the X-ray emitting material
is in fact not perfectly centered on Algol B, but is shifted
slightly inward toward the primary star.  The effective orbital
radius of the X-ray emission allows us to place a constraint on the
possible contribution from Algol A---emission that would skew the
apparent X-ray Doppler radius toward the center of gravity.

As mentioned in \S\ref{s:intro}, it has been argued that 
accretion onto the primary star could be a source of X-rays.  If
this were true, a small X-ray contribution from the accretion activity of
Algol A could explain the inward shift of the effective radius of
X-ray emitting material.  Assuming that the emission of the K star
corona is centered on the star itself, an effective radius of $9.93
R_{\odot}$ for the X-ray emission would indicate that $\sim$85\% of
the total X-ray emission is from Algol B, and the remaining 15\% is
from Algol A.

\citet{Singh1996} considered the issue of accretion-driven X-ray
emission in Algol-type binaries in their comparison between Algols
and RS~CVn-type binaries.  The latter are comprised of two late-type
stars in which neither component filled their Roche lobes.  
\citet{Singh1996} found that the Algol-type binaries are in fact 
slightly X-ray
deficient relative to their RS~CVn cousins, suggesting strongly that
the accretion activity of Algol-type binaries is not a significant source of
X-rays.  The 15\%\ effect we are seeking here would, however, 
be quite inconspicuous in this type of statistical study.  

The possibility of accretion giving rise to significant X-ray flux in
Algol was reviewed by \citet{Pustylnik1995}.  Two key parameters are the
shock temperature of the accreting gas, and the mass transfer rate.
It seems unlikely that the accretion shock can exceed $6\times
10^6$~K, and Pustylnik suggests a maximum X-ray luminosity from
accretion of about $10^{30}$~erg~s$^{-1}$ based on a mass transfer
estimate of order $10^{-11} M_\odot$~yr$^{-1}$.  The X-ray luminosity
of Algol as measured from the HETG observation analyzed here is about
$9\times 10^{30}$~erg~s$^{-1}$ \citep[][in preparation]{Testa2004}; 
accretion does therefore appear sufficient to
account for $\sim 15$\%\ of the total luminosity.  One test of this would be 
whether or not lines formed only above  $6\times 10^6$~K---lines that
could only plausibly originate from coronal emission---show the
same center of gravity as lines formed at cooler temperatures.
Unfortunately, our data are of insufficient quality to perform this
test since only weaker lines in the spectrum exclusively originate
from such hot plasma.  The brightest line formed at these higher
temperatures is H-like Si~XIV $\lambda$6.18.  The line-of-sight velocity as
a function of orbital phase indicated by this line is consistent within
statistical uncertainties with the velocity seen in
the other, cooler lines (Figure~\ref{f:orbit_all}).  However,
measurement errors are sufficiently large that we cannot rule out
differences at the 15\%\ level.

Evidence for plasma associated with accretion and with temperatures of
at least $10^5$~K has, however, been found in the Algol systems
V356~Sgr and TT~Hya from recent FUSE observations 
\citep[][respectively]{Polidan2000,Peters2001}.  Emission detected from
O~VI appeared to be associated with a bipolar flow that makes a large
angle with the orbital plane.  However, in order for this plasma to contribute
significantly at X-ray wavelengths, it must be comprised of components at
least an order of magnitude hotter than the formation temperature of
O~VI.   

\begin{figure} [p]
\epsscale{1}
\plotone{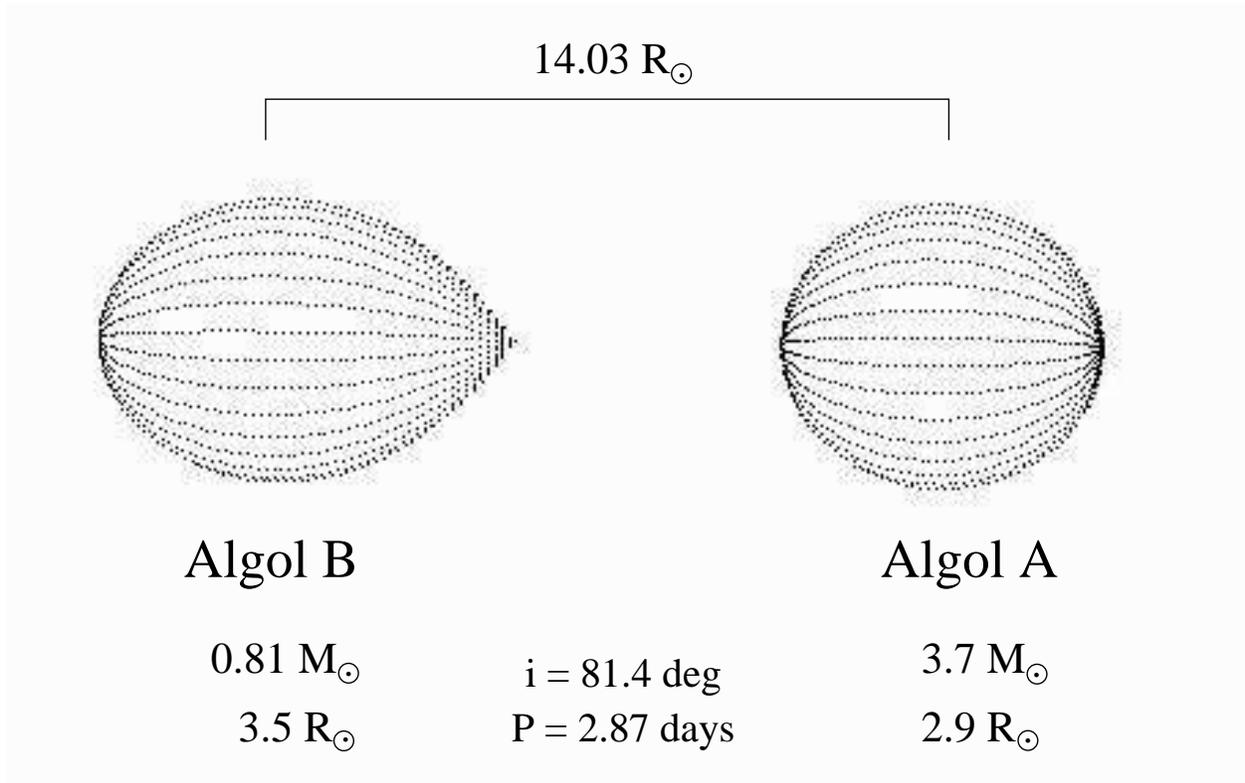}
\caption{A model of Algol at quadrature.  Algol B, which
  nearly fills its Roche lobe, has an extension of material toward the
  system center of mass.  This type of assymetry in the corona of
  Algol B could explain why the effective radius of X-ray emitting
  material is not centered on Algol B, but is slightly
  shifted toward the primary star.}
\label{f:geom}
\end{figure}

Another possible, and we suggest more likely, explanation for the
inward shift of the effective radius is that the corona of Algol B is
not spherically symmetric or exactly centered on the center of mass of
Algol~B, but rather has some asymmetry and structure on the side
facing inward, toward Algol A.  The surface of Algol~B itself will be
severely distended in this direction by the gravitational field of
Algol~A---the Roche lobe filling factor for Algol~B is expected to be
very close to unity, such that any equatorial corona could well lie
beyond the L1 point and would not be gravitationally bound to Algol~B.
We illustrate the approximate geometry of the 
system in Figure~\ref{f:geom}.  This figure was produced using the
{\em Nightfall}
program\footnote{http://www.lsw.uni-heidelberg.de/users/rwichman/Nightfall.html}
by R.~Wichmann, which accounts for the relative masses of the two stars,
the Roche lobe filling factors, and inclination
of the orbit.  

We also note that, in the X-ray lightcurve 
(Figure~\ref{f:lc}), we see a flare just as Algol B comes
out of eclipse.  This flare does not have a sharp rise, but a more
gradual one that appears to start before eclipse egress.  It is
possible that this flaring plasma is located on the hemisphere facing
Algol~A, and that the characteristics of its intensity evolution over
time are modulated by the eclipse.  If the flaring plasma is indeed
located on the hemisphere facing Algol~A, this could also explain why 
the effective radius of the
X-ray emission during this observation is shifted toward Algol A.

\subsection{Non-thermal Broadening}

By comparing observed with theoretical line widths, we have some found 
evidence for moderate excess line widths that we attribute to the
possible presence of 
non-thermal broadening.  While not definitive, 
this result is supported by recent analyses
of far ultraviolet spectra obtained by the Hubble Space Telescope
\citep{Ayres2003} and the Far Ultraviolet Explorer
\citep{Redfield2003}.  Both
studies report the need for excess line broadening to understand the
profiles of the forbidden lines of Fe~XXI at 1354~\AA\ and Fe~XVIII at
976~\AA\ in the more active stars of their
sample with the largest projected rotational velocities ($v\sin i$).
 
\subsubsection{Flows or Explosive Events?}

One source of excess broadening might be turbulence or ``explosive
events''.  Our line profile analysis (e.g.,\ Figures~\ref{f:turbulent}
\& \ref{f:profile}) suggests that a random velocity component of 
$\sim 150$~km~s$^{-1}$ could explain the observed line widths.  This 
is similar to the sound speed at $10^7$~K.  Such a velocity is
reminiscent of the non-thermal broadening of over 100~km~s$^{-1}$ 
seen in transition region lines of stars of different activity as
reviewed by \citet{Wood1997}.  It was suggested in this body of work
that the broadening might be caused by the acceleration of plasma in
magnetic reconnection events associated with microflaring.  It is
possible that the broadening we see in hot X-ray lines is related to
this.  

Other types of flows are also possible.  \citet{Winebarger2002}
observed flows with velocities up to 40~km~s$^{-1}$
in solar coronal loops based on detailed Transition Region and Coronal
Explorer ({\it TRACE}) time-resolved imaging and simultaneous Solar
and Heliospheric Observatory ({\it SoHO}) SUMER spectra.  While the
velocities inferred here are three times this, it is not difficult to
envisage faster flows when considering that the coronal energy
deposition rate of Algol~B is $\sim 10000$ times that of the Sun.

However, as pointed out by \citet{Redfield2003}, in both the above
cases the broadening mechanisms should be ubiquitous among the more
active stars and should not be seen only in those with the largest
$v\sin i$.


\subsubsection{Rotational Broadening and Coronal Scale Height}
\label{s:nonthermal}

Another possible source of excess line widths is rotational broadening
originating in a corona with significantly radially-extended
structure.  As noted in \S\ref{s:intro}, the evidence from
observations of Algol during eclipse does not provide an unambiguous
picture, with some observations requiring apparently extended coronae
\citep{White1986}, and others being consistent with scale heights of a
stellar radius or less \citep{Antunes1994,Ottmann1994}.  

The scale height for plasma of mean molecular weight $\mu$ at a
temperature $T$ on a star of mass $M$ and radius $R$ is $kTR^2/\mu
m_HGM$.  The parameters for Algol~B were summarized recently by
\citet{Drake2003}, who adopted $R_B=3.5R_\odot$ and $M=0.8M_\odot$ 
\citep[see also][]{Richards1988,Kim1989}.  For a coronal temperature $T\sim
10^7$~K on Algol~B, the scale height is then about $1.4R_B$ in the
absence of strong centrifugal forces.  
Our median scale height and 1$\sigma$ upper limit are 3.1$R_{B}$ and 
3.8$R_{B}$, respectively.  
While this is slightly larger than the 
thermal scale height in the absence of strong centrifugal forces
arising from rapid rotation, it is
important to keep in mind the following things.  
Firstly, the line profile
analysis depends on subtraction of the dominant instrumental and
thermal broadening components and is inherently prone to additional
systematic error arising as a result of imperfections in the
description of these dominant effects.  Secondly, for a corona on
Algol~B with a radial extent comparable to the stellar radius, the
competition between gravitational and centrifugal forces becomes
significant: centrifugal acceleration on a single star with the same
rotation rate is equivalent to gravitational acceleration at a height of
$1.2R_B$.  While the binary nature of Algol complicates the picture, it
is possible that some regions of an extended corona comprise plasma
bound by magnetic fields, as has been discussed in connection with
other rapidly rotating stars, such as AB~Dor
\citep[e.g.][]{Jardine1999}.  (See also \citet{Schmitt1997} who
concludes that the 1032~\AA\ O~VI profile of AB~Dor, observed with
ORPHEUS, is not rotationally broadened beyond the photospheric surface
$v\sin i$ value.)

We conclude that, in the interpretation that the tentative evidence
for excess line widths is the result of rotational broadening, the
extension of the corona would be similar to the expected thermal scale
height and comparable to the stellar radius. 
This interpretation is consistent with that of \citet{Redfield2003}
based on their FUSE analysis of active stars.  These authors suggested
that their excess line widths could result from coronal structures
with heights up to $1.3R_B$.  Such a picture would also be
consistent with the eclipse studies of \citet{Antunes1994} and
\citet{Ottmann1994}.  With regard to eclipses, it is important to note
that extended structure at the stellar poles would give rise to no
obvious X-ray eclipse.  However, pole-dominated emission also would
not give rise to strong rotational broadening.  Our results suggest
that, if rotational broadening is the correct interpretation for our
observed line widths, this emission must be distributed around the
star and not concentrated at the poles.

The HETG observation does cover the secondary optical eclipse, and
there is some evidence that we do indeed see some obscuration of the
X-ray corona, despite the complications presented by the flare.  We
have computed theoretical X-ray lightcurves for Algol using a
spherically-symmetric, optically-thin emitting shell coronal model.
The model assumes that Algol A is 100\%\ X-ray dark, and uses the
following physical and orbital parameters: $M_B = 0.81 M_{\odot}$,
$M_A = 3.7 M_{\odot}$, $R_B = 3.5 R_{\odot}$, $R_A = 2.9 R_{\odot}$,
$P = 2.867$ days, and $i = 81.4$ degrees
\citep{Hill1971,Tomkin1978,Richards1988}.  Theoretical lightcurves
were calculated for an array of Algol B coronal scale heights, ranging
from 0.1 to 4.0 stellar radii.  Figure~\ref{f:lc} shows the observed
lightcurve of Algol with three theoretical lightcurves of various
scale heights overplotted.  Also plotted is a similar curve computed
for polar emission restricted to high latitudes corresponding to the
scale height predicted by quasi-static loop models (see below).  
If we assume that the count rate at 
phase $\sim$ 0.5 represents the quiescent level of Algol B, as does
the level toward the end of the observation, we can see
from the overplotted models that scale heights of $\sim 1.0$-$2~R_B$
represent the data reasonably well.  The polar emission model clearly
does not represent the observed eclipse minimum; we discuss this
further below.

A scale height commensurate with the thermal scale height would appear
to clash with the recent interpretation of density-sensitive line
ratios seen in {\it Chandra} LETGS observations of Algol by 
\citet{Ness2002}.
The interpretation of helium-like line ratios in terms of plasma
density is complicated in the case of Algol by the radiation field of
Algol~A, which is sufficiently strong to cause significant radiative
excitation of electrons from the upper level of the forbidden line to
the upper level of the intercombination line through the transition
$2\,^3S\rightarrow 2\,^3P$ in the ions C, N and O.  Nevertheless,
\citet{Ness2002} pointed out that, even with densities of order
$10^{10}$~cm$^{-3}$---similar to those found for other active stars in
the extensive survey of \citet[][in preparation]{Testa2004}---simple 
quasi-static coronal loop model scaling laws
\citep[e.g.][]{Craig1978,Rosner1978,Jordan1980} 
that predict that loop peak temperature $T$,
pressure $P$ and length $L$ are related by $T\propto(PL)^{1/3}$,
suggest maximum loop lengths of order a few $10^{10}$~cm.  This is
only a fraction of the radius of Algol~B.

\subsubsection{$f/i$ Ratio and Coronal Extent}
\label{s:f/i_discussion}
  
As mentioned in \S\ref{s:f/i_analysis}, there was no detectable 
orbital modulation of the O~VII $f/i$ ratio in the Algol LETGS data,
which was observed during orbital phase $\phi=0.75$-1.03.  
\citet{Schmitt2003_XMM} also found
no differences in the $f/i$ ratio of Algol spectra which was extracted
during the quiescent and flaring phases of an XMM-Newton
observation.  

This result has important implications.  If we assume that the corona
of Algol is indeed similar to that of other active stars, such as its
``coronal twin'' HR~1099 \citep{Drake2003}, then we must conclude that {\em
the bulk of the O~VII emission is always illuminated by Algol~A}.  This
can be achieved by having a fairly extended corona with a scale height
$\geq 1R_B$, as our line profile analysis indicates.
Instead, {\it if the O~VII emitting coronal structures are compact
compared to the stellar radius, then they must reside predominantly at
the poles} where the Algol~A UV radiation field is not significantly
shadowed at primary eclipse.  If scale heights are fairly large ($\geq
1R$), the classical quasi-static loop scaling laws must then be quite
inapplicable to these coronal structures.  In the case of compact (by
necessity) quasi-static loops with a plasma density of a few
$10^{10}$~cm$^{-3}$, \citet{Ness2002} estimate a coronal volumetric
filling factor of up to 0.3---unreasonably large for emission
confined to polar regions.  As noted earlier in this section, the
synthetic lightcurve computed for the case of polar emission with a
scale height of $0.1R_B$, similar to that predicted by quasi-static
models, is a very poor match to the observed eclipse minimum light
(Figure~\ref{f:lc}).  This model predicts very sharp eclipses which
have never been observed in X-ray observations of Algol.

In summary, the combined evidence of a tentative detection of 
excess line widths, the shape of the X-ray lightcurve, and the lack of
orbital modulation of the O~VII $f/i$ ratio going into primary
eclipse, all point to the plasma source at temperatures up to
several $10^6$~K having a scale height of the same order as the
stellar radius, and being distributed around the K star.
Such distributed coronal structure could be responsible for, e.g., the
flaring at lower latitudes, which was inferred by
\citet{Schmitt2003_XMM} based on an XMM-Newton X-ray lightcurve.
Unfortunately, the X-ray data alone cannot rule out other scenarios involving
compact structures, though these would require special and fortuitous
placement of dominant active regions in order to remain consistent
with past and present observations of the coronal emission of Algol.
\citet{White1986} drew similar conclusions based on the failure to
observe a strong X-ray eclipse in the EXOSAT observation, though with
the benefit of not having seen data from subsequent observations that have
led to strong ambiguities in interpretation.  Our favored picture of
the X-ray emission of Algol is also consistent with the picture
painted by \citet{Favata2000}, and with the interpretation of excess
line widths seen in UV and FUV coronal forbidden lines by
\citet{Redfield2003}.  What we have not tackled in this
article, however, is the very hot ($\geq 10^7$~K) plasma that often
flares and seems to be located at the poles.  The recent plasma
density survey of \citet[][in preparation]{Testa2004}
indicates that this very hot
material is at much higher densities of $10^{12}$~cm$^{-3}$, and so
occupies quite different structures.  These are likely to be the
structures responsible for pole-dominated flaring activity, such as
that highlighted by \citet{Schmitt1999}.


\section{Conclusions}
\label{s:conclusions}

A detailed analysis of high quality {\it Chandra} HETGS spectra of the Algol
system provides new insights into the geometry of the X-ray emission
in this system.  Based on this analysis we draw the following
conclusions:

\begin{enumerate}

\item Our study clearly reveals Doppler shifts of the X-ray emitting
plasma corresponding to orbital motion with a velocity at quadrature
of 150~km~s$^{-1}$.  These data thus provide the first definitive proof
that the coronal activity of Algol~B dominates the X-ray emission of
the system, as has long been suspected \citep[e.g.,][]{White1980}.

\item The observed Doppler motion of the X-ray plasma on Algol~B appears to
be off-center relative to the stellar center of mass, and shifted
toward Algol~A.  We suggest that this occurs as a result of the tidal
distortion of the surface of Algol~B.  The HETGS light curve exhibits
a flare near secondary eclipse which, if located on the hemisphere
facing the primary B8 star as expected, might also suggest the presence of a
dominant active region that could bias the emission toward a slightly smaller
orbital radius.

\item Alternatively, X-ray activity of Algol~A, possibly as a result
of accretion from Algol~B, could be responsible for the smaller apparent
orbital radius of the X-ray emission.  In this case, we estimate that
such a contribution amounts to no more than 15\% of the total
emission.  Such an X-ray flux would be consistent with current
ideas as to the plausible range of accretion activity in Algol.

\item We have found some evidence for excess line broadening in bright X-ray
emission lines, above that expected from surface rotation and thermal
motions.  If this effect is real, it suggests broadening by rotation of a
radially-extended corona.  The observed widths would require a coronal scale
height of at least one stellar radius.  This would be consistent with
recent observations that detected excess broadening in the forbidden Fe~XVIII
974~\AA\ line, seen in FUSE FUV spectra of rapidly rotating active
stars.

\item While turbulence, microflares or flows could also produce excess
line broadening, the idea of a coronal plasma with significant radial
extent is also supported by UV-sensitive lines of O~VII seen in an
LETGS observation of Algol.  No change in the UV-sensitive lines was
seen going into primary eclipse, when the majority of the visible
hemisphere of Algol~B was shadowed from the radiation field of
Algol~A.

\item While it is possible that other physical effects are responsible
for excess line widths and lack of modulation of the UV-sensitive
O~VII lines, coronal extension would appear to be the most likely.  A
consistently high plasma density of a few $10^{11}$~cm$^{-3}$, as
would be indicated by the O~VII lines in the absence of UV excitation,
would be unlike {\em any} of the densities seen at O~VII
temperatures in the sample of active stars surveyed by \citet[][in
preparation]{Testa2004}.  This picture also appears to be consistent
with conclusions drawn from recent lightcurve analyses which were
summarized in \S\ref{s:intro}.

Coupled with the results of the coronal density survey of \citet[][in
preparation]{Testa2004}, we can summarize as follows.  Coronal plasma
with temperatures of up to several $10^6$~K appears to have a
significant component lying in extended structures with a scale size
similar to the stellar radius and densities typically of a few
$10^{10}$~cm$^{-3}$---similar to solar active regions.  Very hot
coronal plasma appears more exclusively in much more compact regions
with densities of order $10^{12}$~cm$^{-3}$, and in very active,
rapidly rotating stars appears to be concentrated more toward stellar
poles.  Further spectroscopic studies of these apparently different
coronal regimes, especially with regard to possible abundance anomaly
differences between them, should prove very interesting.

\end{enumerate}

This work was supported by NASA contracts NAS8-39073 and
NAS8-03060 to the {\it Chandra} X-Ray Center, as well as NAG5-9322 which
provides financial assistance in the development of the PINTofALE
package.


\clearpage
\begin{table}  [h]
\begin{center}
\caption{The observed and predicted line widths of
  coadded Algol and Capella data, and the excess line
  widths detected in Algol.  The upper portion shows line widths for
  MEG data, and the lower portion shows line widths for HEG data.
  All line widths are given in units of \AA.}

\normalsize

\vspace{0.5in}
\begin{tabular} {c c c c c c}

\hline
\vspace{-0.1in}

 & & & 
\multicolumn{1}{c}{Algol} & \multicolumn{1}{c}{Capella} \\

 $\lambda_{o}$ & Algol observed &  Capella observed & 
 predicted & predicted & Algol excess \\
\hline

24.78 & 0.0306 $\pm$ 0.0033 & 0.0217 $\pm$ 0.0055 & 0.0285 & 0.0267 &  0.0021 \\
18.97 & 0.0256 $\pm$ 0.0013 & 0.0216 $\pm$ 0.0009 & 0.0249 & 0.0238 &  0.0006 \\
16.78 & 0.0284 $\pm$ 0.0025 & 0.0190 $\pm$ 0.0007 & 0.0212 & 0.0211 &  0.0072 \\
15.01 & 0.0266 $\pm$ 0.0017 & 0.0196 $\pm$ 0.0005 & 0.0199 & 0.0199 &  0.0067 \\
12.13 & 0.0266 $\pm$ 0.0008 & 0.0205 $\pm$ 0.0006 & 0.0217 & 0.0212 &  0.0049 \\
 8.42 & 0.0211 $\pm$ 0.0009 & 0.0191 $\pm$ 0.0008 & 0.0197 & 0.0195 &  0.0014 \\

\hline
\hline

12.13 & 0.0170 $\pm$ 0.0008 & 0.0164 $\pm$ 0.0008 & 0.0146 & 0.0140 &  0.0024 \\
 8.42 & 0.0127 $\pm$ 0.0008 & 0.0139 $\pm$ 0.0010 & 0.0134 & 0.0130 & -0.0006 \\

\hline
\label{t:linewidths}
\end{tabular}
\end{center}
\end{table}

\clearpage

\end{document}